\documentclass[12pt]{article}

\usepackage{enumitem}
\usepackage[scaled]{helvet}
\usepackage{stix2}
\usepackage[scaled=0.9]{zi4}
\usepackage{xcolor}
\usepackage[T1]{fontenc}

\usepackage{titlesec}

\titleformat{\section}{\normalfont\sffamily\large\bfseries\color{black}}{\thesection.}{0.3em}{}
\titleformat{\subsection}{\normalfont\sffamily\small\bfseries\color{black}}{\thesubsection.}{0.3em}{}

\usepackage[onehalfspacing]{setspace}
\usepackage{authblk}
\usepackage{geometry}
\usepackage{caption}
\AtBeginDocument{\frenchspacing}
\usepackage{microtype}

\usepackage{standalone}
\usepackage{tikz}
\tikzset{font=\fontfamily{helvet}\selectfont}
\usepackage[noabbrev]{cleveref}
\usetikzlibrary{patterns,arrows,arrows.meta,calc,shapes,shadows,decorations.pathmorphing,decorations.pathreplacing,automata,shapes.multipart,positioning,shapes.geometric,fit,circuits,trees,shapes.gates.logic.US,fit}
\usetikzlibrary{backgrounds,scopes}

\usepackage{url}
\usepackage[style=ieee]{biblatex}
\begin{document}
\title{\Large Navigating the sociotechnical labyrinth: Dynamic~certification~for~responsible~embodied~AI}
\author[1]{Georgios Bakirtzis}
\author[1]{Andrea Aler Tubella}
\author[1]{Andreas Theodorou}
\author[2]{David Danks}
\author[3]{Ufuk Topcu}
\affil[1]{Universitat Politècnica de Catalunya}
\affil[2]{University of California, San Diego}
\affil[3]{The University of Texas at Austin}

\date{}
\maketitle
\begin{abstract}
Sociotechnical requirements shape the governance of artificially intelligent~(AI) systems. In an era where embodied AI technologies are rapidly reshaping various facets of contemporary society, their inherent dynamic adaptability presents a unique blend of opportunities and challenges. Traditional regulatory mechanisms, often designed for static---or slower-paced---technologies, find themselves at a crossroads when faced with the fluid and evolving nature of AI systems. Moreover, typical problems in AI, for example, the frequent opacity and unpredictability of the behaviour of the systems, add additional sociotechnical challenges.
 
To address these interconnected issues, we introduce the concept of \emph{dynamic certification}, an adaptive regulatory framework specifically crafted to keep pace with the continuous evolution of AI systems. The complexity of these challenges requires common progress in multiple domains: technical, socio-governmental, and regulatory. Our proposed transdisciplinary approach is designed to ensure the safe, ethical, and practical deployment of AI systems, aligning them bidirectionally with the real-world contexts in which they operate. By doing so, we aim to bridge the gap between rapid technological advancement and effective regulatory oversight, ensuring that AI systems not only achieve their intended goals but also adhere to ethical standards and societal values.
\end{abstract}

\section{Introduction}

Autonomous technologies are revolutionising contemporary society, calling into question traditional policies and modes of regulation~\cite{danks:2017,floridi:2018,mokander:2022}. The primary driver for this paradigm shift is artificial intelligence (AI), which underlies systems that can demonstrate highly context-sensitive and emergent behaviors. In particular, AI systems are able to do inferences, decisions, and actions that have historically been thought to require human cognition.\footnote{There are many debates about the exact definition of AI. We sidestep those issues here, as one virtue of a dynamic certification is that it does not depend on any particular characterization of AI.} While these systems' flexibility and creativity provide compelling reasons to adopt AI over conventionally engineered systems for self-sufficiency, they also present unprecedented challenges to our ability to design and implement \emph{trustworthy}~\cite{neumann:2022}, \emph{safe}~\cite{khlaaf:2023}, and ultimately \emph{well-calibrated} autonomous systems~\cite{uspresident:2019}. On the one hand, the potential for AI to adapt, learn, and respond in dynamic ways is what sets it apart from traditional technologies. On the other hand, this adaptability makes it challenging to predict or understand the performance of these systems. As a result, comprehensive and effective governance measures have proven to be a complex and often daunting goal.

Regulatory and certification bodies across the globe have been grappling with these issues. The European Union (EU) has, for example, engaged in extensive efforts to align growth and innovation in AI with societal values and legal norms~\cite{eu:2019}. The recently voted ``AI Act'' is meant to establish a cross-domain minimum set of requirements of AI systems by following a risk-based approach \cite{eu:2024}.
In the United States, different agencies have begun to expand their expertise to account for autonomous systems and the ways that they are covered by existing rules and regulations. Broader U.S. efforts---for example, the Blueprint for an AI Bill of Rights~\cite{wh:aibillofrights} or the NIST AI Risk Management Framework (RMF)---have aimed to provide overarching frameworks to support and address various aspects of AI governance.

These regulations complement the over $700$ guidelines and other soft policy documents released over the past decade around the world which provide recommendations (sometimes quite abstract) around the responsible use of AI technologies~\cite{theodorou:2020}. There is significant variability within these approaches for governing this fast-changing technology. Some approaches opt for focusing on the application domain for the AI system, with different rules for medical applications, than for entertainment, for example. Other approaches focus on the type of technology, for example distinguishing facial analysis systems from recommender systems, with very different rules applying to each. Still other approaches focus on the level and type of risk,  and setting stricter rules for those uses and contexts considered potentially more risky.
Despite this variability, some fundamental values are found across all legislation and recommendations, such as: transparency, safety, privacy, and non-discrimination~\cite{jobin:2019}. Ensuring that AI systems adhere to these values is central to responsible AI governance.
However, the interpretation of those values---and, hence, means of achieving them---differs between societies and even individual actors in those societies~\cite{theodorou:2020,aler:2019}.

Once rules and regulations are set, there are additional challenges for those developing AI systems to assess, ensure, and validate that a system adheres to those rules.
In particular, multiple gaps remain in our ability to audit AI effectively and to check for compliance with regulations. Many AI systems operate as ``black boxes,'' where the processes leading to a particular output or decision are opaque and too complex to scrutinize, raising vital concerns about accountability, ethics, and public trust~\cite{acemoglu:2021}. Essential aspects of development and deployment are rarely captured in documentation or metadata. We often lack an understanding of a system's normal operating conditions and the situations when it may misbehave. Moreover, the rapid pace of AI development often outstrips the speed at which regulations can be formulated and implemented, leading to confusion about which regulations apply and how.

\begin{figure}[!t]
    \centering
    \includegraphics[width=.7\linewidth]{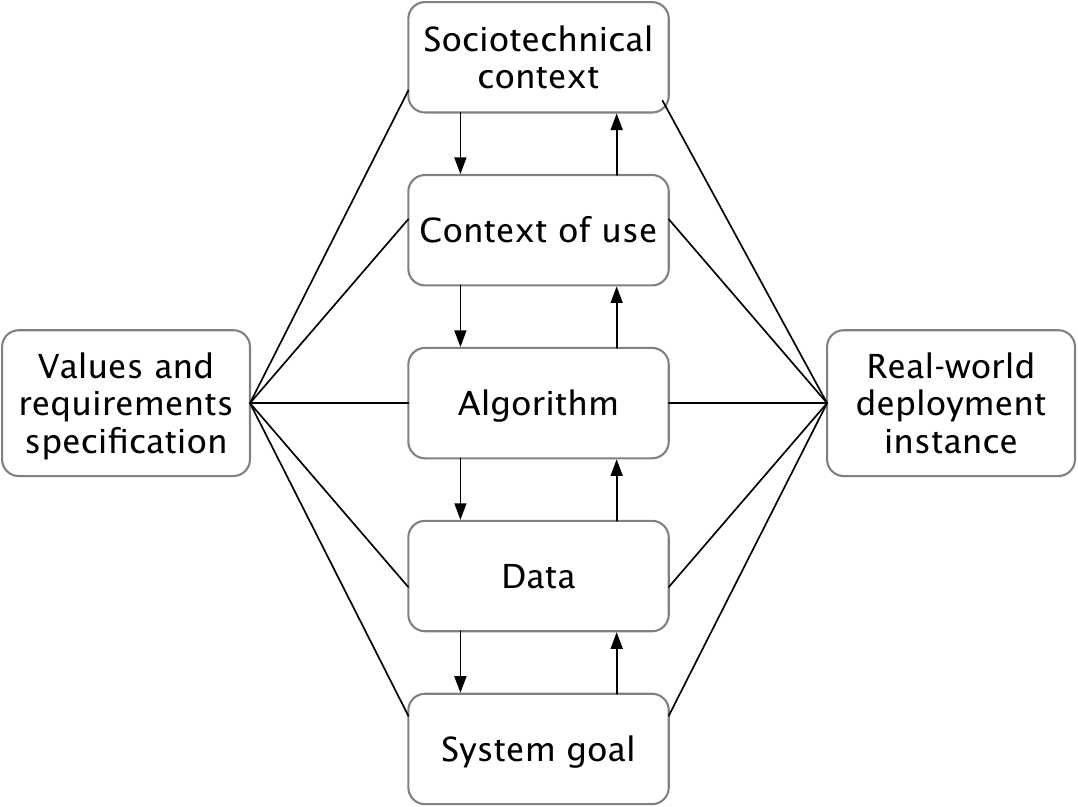}
    \caption{The dynamic certification for sociotechnical AI systems contains bidirectional, interdependent and layered challenges.}
    \label{fig:challenges}
\end{figure}

Overall, navigation of evolving AI regulation demands a multifaceted approach that focuses on several layers:
\begin{itemize}
    \item[i)] Sociotechnical context: The sociotechnical context determines which requirements apply, as well as how they are interpreted in each situation by each stakeholder.
    \item [ii)] Context of use: The context (or contexts) of use determine which requirements must be followed and why, depending on the specific description of the situation (its criticality, which stakeholders are interacting with the system, etc.)
    \item [iii)] System: The specific workings of the system being regulated sometimes must be directly examined and tested for compliance.
  \end{itemize}

Simultaneously addressing all of these aspects requires a layered framework that can provide a structured guide to ensure compliance of AI systems with the complex requirements spanning ethical, legal, and functional domains (figure~\ref{fig:challenges}). Within this framework, each layer presents its own set of challenges, such as: capturing diverse and sometimes conflicting requirements, ensuring use cases and simulations adapt to real-world scenarios~\cite{hofer:2021}, aligning computational models with ethical or legal guidelines~\cite{ehsan:2022}, representing mutable social contexts, appropriately acquiring and labelling data~\cite{liang:2022}, and formalising dynamic legal and ethical rules. These challenges intersect and influence one another. For example, limitations in algorithmic modelling may impact what use cases can be fulfilled, which in turn could conflict with societal or regulatory expectations. 

In this chapter, we propose a certification method deriving from systems and their context in such a way that stakeholder-led sociotechnical requirements (legal, ethical, context-specific) can be specified and the system can be checked for adherence. We do so by leveraging the tools of dynamic certification~\cite{bakirtzis:2022}, as these are particularly well suited to this set of challenges. In general, dynamic certification involves iterative development of a regulation or test for specific uses in particular contexts, followed by information-gathering about system performance for ``nearby'' uses or contexts, leading (as appropriate) to expansions in the permissible uses or contexts. This type of approach has been widely used in medical domains, including certification of novel pharmeceutical interventions, which are usually approved for only narrow uses and contexts, and then gradually extended as safety is established. Dynamic certification is best for situations where guarantees---whether safety, performance, or other---are not available, but we nonetheless need to ensure that systems behave appropriately. It is thus ideal for autonomous, embodied AI systems.

\section{AI governance challenges: Towards dynamic frameworks}

The first key aspect when it comes to AI governance is the context-specificity of requirements, especially those relating to law and ethics.
Different requirements may apply to different contexts.
Even cross-border legislation initiatives---for example, a directive from the European Union---may have different local interpretations and, hence, requirements \cite{custers:2018} depending on the complex interplay between national and international laws, as well as differing national priorities and policies.

In addition, when it comes to AI, the vastness of application domains and technical techniques make the development of a single ``horizontal'' regulation or law (that is, applicable across all domains) extremely difficult to execute.
The AI Act, the first legally-binding horizontal regulation on AI, takes a risk-based approach \cite{ruschemeier:2024} that applies different regulatory requirements---perhaps an outright ban on use within the EU---depending on initial and ongoing risk assessments.
However, in the AI Act, the necessary conformity assessment requires grounding and connection with a plethora of other sector-specific rules, regulations, and laws; that is, in practice, the EU AI Act has a significant vertical (that is, sector-specific) component, even though it is written as a horizontal regulation. 
In fact, this latter feature creates conflict between the Act's requirements, set through newly implemented standards, and government standards set by public agencies \cite{ruschemeier:2024}.
Such conflicts need to be resolved if the Act is to be successfully implemented.

One solution would be to turn towards ad-hoc rule-making, in which a regulatory or oversight body takes a largely case-by-case approach to governing AI systems, rather than defining rules and standards across the board. 
For example, within the US, the Consumer Financial Protection Bureau (CFPB) has pursued this strategy through rulemaking by enforcement actions. Internationally, negotiations within groups such as the Technology and Trade Council (TTC) may help to reconcile governance approaches in sector-specific, though ad hoc, ways. These approaches can flexibly adapt to novel technologies but risk inconsistent interpretations and applications between sectors, leaving stakeholders with risk uncertainties. The net result is likely increased costs of developing and deploying AI systems for real-world environments, with all of their sociotechnical complexity.\footnote{Similar issues have bedevilled privacy regulation, with substantial uncertainties about which governance mechanisms are sufficient for requirements of other jurisdictions (for example, does a system that satisfies the California Consumer Privacy Act requirements most likely also satisfies the EU General Data Protection Regulation).%
} Thus, we
require a dynamic regulatory and governance approach, that can capture the case-by-case context-specific legal requirements in a systematic way.

Ethical principles and guidelines are similarly fractured~\cite{theodorou:2020,mittelstadt:2019}: AI ethics, unlike medical ethics, lacks a homogeneous history, professional culture, identity, or extensively developed professional ethics framework. Moreover, the AI development community comprises individuals and teams from varied disciplines with different historical, cultural, and moral backgrounds. Its closest analogue--software engineering--similarly lacks a shared history of clear professional ethics, including fiduciary duties to the public. Put simply; there is no widely accepted standard for being an ethical AI developer or software engineer. In addition, AI's impact can be far-reaching and not immediately apparent to developers, which increases the risk of unintentionally unethical professional behavior. Ethical risks in AI are continuous and not always experienced directly by those affected by the system, further complicating the creation of standards for robust and trustworthy AI development. Indeed, it is acknowledged that concepts such as trustworthiness, explainability or fairness highly depend on the context and use-case~\cite{liao:2022}, with stakeholder involvement needed to ascertain what each of these aspects means in context~\cite{delgado:2023}. This means that, often, highly context-specific requirements arise.

AI governance is further complicated by the need for regulations to be robust to the dynamic nature of AI technologies and their deployment contexts. AI systems constantly learn and evolve, as does our understanding of their capabilities and limitations. Static regulatory frameworks (for example, one-time certifications or fixed performance standards) will thus fall short in addressing new challenges or ethical issues arising from advancements in AI technologies or deployments. Instead, we require a dynamic regulatory and governance approach.

In this vein, a range of high-level goals have been proposed for better regulating and governing AI systems, such as the need for standards to be flexible, regularly updated, applicable across multiple sectors, and tuned to intended scopes or uses. More generally, there is increasing recognition that regulation will need to shift from a purely reactive to a more proactive mode, enabling innovation while remaining grounded in preexisting standards for system assurance. And, of course, the distribution of standards should be open, transparent, inclusive, accessible, and consensus-based~\cite{harvey:2020, nist:2019}. Finally, we must recognize that it is no longer possible to provide a complete delineation of concerns before development; instead, multi-stakeholder requirements must be considered before design time, during design time, and even after deployment~\cite{falco:2021}.

In the face of these many challenges, we propose that AI regulation should involve a multi-pronged approach that incorporates diverse stakeholder requirements within a flexible regulatory framework containing continual feedback loops as systems evolve and learn ``in the wild.'' This approach holds the promise of appropriate guidance in the public interest without stifling technological innovation. Of course, we are not the first to suggest such a framework; the critical challenge is implementing such a dynamic certification system.

\section{Dynamic certification}

As we continue to develop and deploy autonomous systems, we are increasingly recognizing the context-dependent nature of their performance. These systems---whether autonomous vehicles or drones, or robotic systems---exhibit varying levels of competence and reliability depending on the specific environments and scenarios in which they operate. This realization has exposed the limitations of traditional static certification frameworks, which typically assess a system's capabilities at a single point in time and under a narrow set of predefined conditions.

Static certification approaches, while valuable for many conventional technologies, prove inadequate for the dynamic and evolving nature of AI-driven autonomous systems. These systems learn and adapt over time, potentially expanding their capabilities or developing unforeseen behaviors as they encounter new situations.
In fact, any system as it interacts with the environment will output emerging behavior that is hard to predict and understand \cite{theodorou:2017}.
Consequently, a one-time approval process fails to capture the ongoing changes in system performance and the associated risks.
The challenge, therefore, lies in implementing a process that can adapt to the evolving capabilities of these systems while continuously ensuring safety and reliability.

These considerations naturally suggest a turn to dynamic certification, an iterative approach that allows for the ongoing assessment and adjustment of what autonomous systems should and should not be allowed to do in various operational contexts. Dynamic certification offers a structured approach for regulating autonomous AI systems that adapts to their evolving capabilities. This framework allows for the gradual expansion of a system's approved operational envelope based on demonstrated performance and safety in real-world conditions. Rather than providing an one-time approval, dynamic certification implements a continuous evaluation process that responds to the system's actual behavior and learning over time.

The core principle of dynamic certification is the incremental increase in autonomy and operational scope as we gain evidence and knowledge. Initially, an autonomous system might be certified for limited operations under strict constraints. As the system proves its reliability and safety within these boundaries, its certified operational envelope can be expanded, certifying more complex behaviors. This might involve allowing the system to operate in more diverse environments, handle more complex tasks, or make higher-level decisions with less human oversight. This approach involves continuous monitoring of the system's performance in real-world environments, gathering comprehensive data on how the system behaves across a wide range of scenarios, including edge cases that may not have been anticipated during initial testing.
As the system demonstrates consistent reliability and safety within its current operational parameters, dynamic certification provides a framework for gradually increasing its autonomy. This incremental expansion of capabilities is always based on both empirical evidence of the system's performance and theoretical projections. The certification adjustments are made through rigorous ongoing assessments, which evaluate not only the system's technical performance but also its interactions with humans and other systems in its operational environment.

This approach is particularly relevant for embodied systems. For instance, an autonomous drone system might initially be certified for operations in rural areas during daylight hours. As it demonstrates consistent performance and safety, its certification could be expanded to include operations in more densely populated areas or during nighttime, always contingent on ongoing monitoring and performance evaluation.

Additionally, dynamic certification also incorporates mechanisms for restricting or revoking permissions if a system's performance degrades or if new risks are identified. This bidirectional flexibility allows overseers to respond promptly to emerging safety concerns while also providing a pathway for systems to expand their capabilities as they improve.

Implementing dynamic certification effectively requires a combination of formal methods and precise technical knowledge. This includes developing robust metrics for measuring system performance, creating verifiable safety criteria, and establishing clear thresholds for expanding or restricting operational envelopes. It also necessitates the use of advanced monitoring tools and data analysis techniques to process the large volumes of performance data generated during real-world operations.

Moreover, dynamic certification can help address the ethical challenges of fairness and non-discrimination in AI deployment. By continuously monitoring and evaluating system performance across diverse scenarios and populations, it can help identify and mitigate potential biases or unintended consequences that may emerge over time.

By focusing on context-specific approvals and real-time monitoring, dynamic certification aims to balance innovation with safety, allowing for responsible deployment and expansion of autonomous AI capabilities during deployment.

\section{Dynamic certification in action}

\begin{figure*}[!t]
    \centering
    \includegraphics[width=\textwidth]{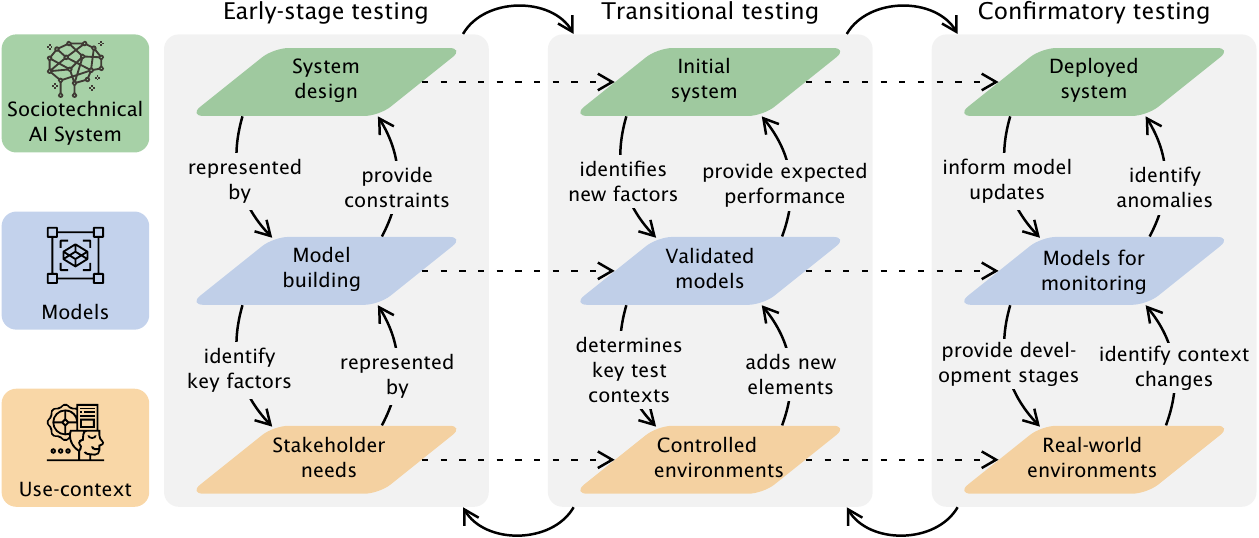}
    \caption{Dynamic certification explicitly allows for ambiguities in the initial specifications, uncertainties, and decisions yet to be made. In addition, dynamic certification keeps a decision trail by intertwining modeling and testing throughout the lifecycle. Models are not only represented by code, but also living documents continually updated in response to changing data and play a useful role in recording changing assumptions and specifications. Models, specifications, and tests are continually refined as we better understand real-world contexts.}
    \label{fig:dynamic-certification}
\end{figure*}
Consider the design and deployment of an advanced delivery robot for urban areas. This is a complex task: it can be challenging to have an effective delivery robot even in an austere environment, but cities are bustling, chaotic, and constantly changing. How can we ensure our robot safely and effectively carries out its duties, including delivery and behaviors from dodging pedestrians to navigating traffic? While a testbed or lab environment might provide valuable insights, it is implausible that we could have a set of regulatory standards that apply to all urban environments for useful lengths of time. We show how our approach (\cref{fig:dynamic-certification}) would provide a precise, adaptable approach to dynamic certification for this type of system.

The process begins with specifying a ``formal model''--essentially, a draft blueprint for the robot's physical design and required behaviors in various situations.%
These formal models enable designers to anticipate challenges and avoid potential design clashes precisely, and so are critical early in the system's lifecycle. For instance, while our delivery robot should be fast to meet delivery times, it also should not speed in crowded areas. Recognizing and addressing these conflicting demands early on saves time, money, and potentially, mishaps. 

As we refine these formal models and establish precise guidelines, we move to virtual simulations to test our robot's behavior. As simulated agents navigate virtual worlds, the robot should be tested in high-fidelity simulators based on real-world physics. This ``early-stage testing'' involves the iterative refinement of the system design and desiderata, the formal models of the system, and the critical needs and goals of diverse stakeholder groups.%
The aim is to establish that it is highly likely that the AI system will essentially perform as required, though with the recognition that no simulator can be perfect or complete.

Thus, we must move to ``transitional testing,'' which involves targeted testing of the AI system in controlled, real-world environments, such as an urban-like special-built environment or a secluded part of an actual city.\footnote{The Virginia Department of Transportation race track for cars is an example of this type of controlled environment, albeit for autonomous vehicles.} This phase introduces the complexity of the real world but in controlled ways that enable us to fine-tune the system's operations, ensuring its behavior aligns with our formal models and real-world demands. Importantly, this phase requires the deliberate construction of scenarios for which the AI system is likely to \emph{fail}. We are not simply trying to meet some benchmark but rather to use the controlled environment to improve and refine the (validated) formal models and the AI system itself.

Once the robot's abilities have been certified in testing scenarios, we can move into actual urban settings with ``confirmatory testing.'' As with transitional testing, this phase is not simply a free-for-all or ``deploy and forget'' approach. The robot must be closely monitored so the data we gather can help refine our models and our understanding of its operations in real-world scenarios.\footnote{This phase is thus analogous to the post-market monitoring of drugs and other medical innovations to detect unexpected patterns of adverse events.} We expect that the AI system will (and should) change, evolve, and adapt over time as more is learned about its performance capabilities in diverse environments. Static certification approaches struggle with this type of change; dynamic certification incorporates these updates as a desired feature, not a bug.

Dynamic certification, in its essence, is a response to the evolving nature of AI technology and its interactions with complex real-world environments. 
However, the task of continually adapting this framework to align with rapid technological advancements and changing societal values is a non-trivial endeavor.

\section{Open problems in dynamic certification for embodied AI}

An essential set of challenges for adapting dynamic certification to embodied AI is technical. AI system complexity and a lack of understanding about appropriate uses and failure modes already too often lead to systems engaging in hazardous behavior \cite{consumerreports:tesla, guardian:uber}. Dynamic certification can reduce such occurrences only if engineers, computer scientists, and AI ethicists can provide regulators and stakeholders with the necessary predictive and analytic capabilities within a dynamic certification regime. ``Deploy and monitor'' is a recipe for repeated failures and harmful incidents with AI systems. However, doing better through frameworks of dynamic certification will require substantial advances in methods and uses of predictive and environmental modeling.

We propose that much of the critical research for practical dynamic certification falls under the scope of model-based design. Technically, the relationships between models and reality must become more traceable and computationally tractable. Simulation capabilities must also be augmented with domain- and application-relevant dynamics of the world, such as pedestrian behavior, to enable more precise design and transitional testing. Another challenge is integrating different programming types, marrying AI's adaptability with precise rule-based commands; neurosymbolic programming is one example of such an integration. These potential advances could enable us to synthesize architectures of controllers that provide guarantees of trustworthy, value-supporting behavior.\footnote{Logical systems can provide a helpful language for specifying value structures and connecting those with observable behaviors. At the same time, different types of formal systems can be used to provably verify different properties, such as temporal and resource allocation properties being both specifiable and verifiable (or counterexamples provided) within different types of logical systems.} For early-phase testing, we need advanced simulations that mirror real-world unpredictability. We need formal languages and frameworks to represent the relevant aspects of environments for the other testing phases. 

More generally, as our models and systems evolve to capture critical design changes and adjust to new expected behaviors, we must look beyond the hardware and software of the robot in isolation. For example, while a coder can instruct the robot to avoid a physical obstacle, appropriate interaction with a person trying to retrieve a package requires insights from human behavior experts. Each stage of this process will thus inevitably involve significant sociotechnical research that integrates insights from technology experts with those from social scientists, ethicists, and other stakeholders. However, there is still limited research on best practices for multidisciplinary, cross-functional teams.

As just one simple example, the design and development of an AI system inevitably involve many living documents that incorporate rationales and changes in requirements and expected behaviors through the lifecycle of that system. Sociotechnical research is required on how best to represent and use these decisions to iteratively improve our designs iteratively, particularly in collaboration with key stakeholders. In particular, these decisions should be guided by the community, ethical, and psychological values the AI system should implement, but those values can rarely be expressed purely technically. 

Finally, regulatory challenges loom, even if the sociotechnical and collaborative problems can be solved. We need significant translational research to determine how to generate better testing contexts and standards for AI systems in light of our formal models. Comprehensive testing regimes create challenges for AI design and complicate development, but it is necessary to assure stakeholders that the system will not transition to a hazardous state. In addition, our characterization of dynamic certification has assumed that regulators or other independent experts can recognize and enforce ``good'' system behavior. However, regulatory standards may be set by laws that fail to capture the complexity of AI, or the standards may have been written in ways that do not provide operational precision. Dynamic certification depends on standards tailored for AI systems. Research on ethics, public policy, and their implementation in law and regulation will thus also be required to realize the promise of dynamic certification. 

Dynamic certification offers a path towards safe, ethical governance of AI systems but requires an ambitious multidisciplinary research agenda spanning from computer science to sociology to ethics to public policy. As we develop and refine AI systems like our urban delivery robot, it is crucial to maintain a dynamic, iterative, and multidisciplinary approach. Only then can we ensure such systems' reliable, safe, and efficient operation in real-world environments. The central open problems in embodied AI are in the sociotechnical perspective of system design and testing.

\section{Conclusion}
Early in the system's lifecycle, formal models\footnote{We emphasize that there are potentially many applicable formal models. Our proposal is mainly agnostic about the exact formal modeling framework.} represents information about the AI system's likely use and deployment contexts, informing stakeholders of potentially clashing requirements and design specifications. Once requirements and design decisions are reconciled in this \emph{early-stage testing}, a series of high-fidelity simulators (for example, based on the underlying physics) can be constructed and used to identify likely behavior violations in core deployment contexts. Those violations can then be mitigated or addressed through design changes prior to substantial (and costly) development. In the \emph{transitional testing} stage prior to deployment, controlled testing procedures can be designed to ensure that artificial test and evaluation settings are realistic given current knowledge and expectations, including appropriate variation in potential use and deployment contexts. %
Once appropriate contexts and uses are identified and validated, \emph{confirmatory testing} processes enable deployment of the AI system in those contexts with necessary supervision, ensuring safety while allowing for additional learning that can iteratively improve the formal models, therefore increasing our understanding of what we do and do not know, and guide further testing, therefore increasing our trust in the deployed AI system. 

We contend that dynamic certification provides a regulatory approach with the features and capabilities needed to govern AI and autonomous systems appropriately. However, we require significant advances in theories, methods, and tools to make dynamic certification a practical possibility for regulators and policy-makers. Dynamic certification requires the ability to measure and understand the uncertainty entailed by AI systems; the ability to understand precisely what the models, simulations, and testing vectors mean at each step; the ability to represent appropriate contextual factors; the ability to determine whether values are appropriately implemented in a system; and much more. Absent these capabilities, dynamic certification risks becoming a mere regulatory checklist replicating the current ``deploy and monitor'' types of regulation that are known to be insufficient for the complexity and corner cases introduced by AI systems. There are also bidirectional effects. Initially, regulation may be inadequate to prevent catastrophic losses, requiring investigations, lessons learned, and continual improvements on the compliance requirements of systems and expectations of regulatory bodies.

More generally, the unique challenges posed by the rapidly evolving AI technologies, such as their context-dependent behaviors and far-reaching impacts, necessitate a shift from traditional regulation to a flexible, values-based approach. Existing static regulatory systems and performance-based metrics fall short of addressing the complexities of AI technologies. To navigate these challenges, we advocate for a paradigm shift towards dynamic certification, which emphasizes the importance of underlying ethical values that guide AI behavior and proposes a responsive approach to regulation that evolves with the development of AI technologies.
However, turning theory into practice requires substantial advancements in multiple fields. %
Despite these hurdles, dynamic certification offers a promising pathway and research trajectory toward AI systems' ethical and safe governance, deserving of further exploration and serious consideration.

\printbibliography

\end{document}